\documentclass[twocolumn,prd,aps,preprintnumbers,fleqn,nofootinbib,superscriptaddress]{revtex4}

\usepackage{epsfig}
\usepackage{bm}
\usepackage{amssymb}
\usepackage{amsmath}
\usepackage{color}

\usepackage{color}
\usepackage{babel}
\usepackage{float}
\usepackage{mathrsfs}
\usepackage{amsmath}
\usepackage{cancel}
\usepackage{graphicx}
\usepackage{setspace}

\usepackage{slashed}
\usepackage{simplewick}
\usepackage{ulem}

\begin{document}

\author{Yoshitaka Hatta}
\affiliation{Physics Department, Building 510A, Brookhaven National Laboratory, Upton, NY 11973, USA}
\affiliation{RIKEN BNL Research Center,  Brookhaven National Laboratory, Upton, NY 11973, USA}

\author{Jian Zhou}
\affiliation{\normalsize\it  Key Laboratory of Particle Physics and
Particle Irradiation (MOE),Institute of Frontier and
Interdisciplinary Science, Shandong University (QingDao), 
266237, China }

\title{Small-$x$ evolution of the gluon GPD $E_g$ }

\begin{abstract}
We study the small-$x$ evolution equation for the  gluon generalized parton distribution (GPD) $E_g$ of the nucleon. 
It is shown that $E_g$ at vanishing skewness exhibits  the Regge behavior  identical to the BFKL Pomeron despite its association with nucleon helicity-flip processes. We also consider the effect of gluon saturation and demonstrate that  $E_g$ gets saturated in the same way as its  helicity-nonflip counterpart $H_g$.   Our result has a direct impact on the modeling of $E_g$ as well as the small-$x$ contribution to nucleon spin sum rules. 
\end{abstract}
\maketitle

{\it Introduction.}
With the Electron-Ion Colliders (EIC) in the U.S. \cite{AbdulKhalek:2021gbh} and in China \cite{Anderle:2021wcy} on the horizon, the study of  the generalized parton distributions (GPDs) \cite{Diehl:2003ny} has gained further momentum for its unique role in imaging the multi-dimensional structure of the nucleon. Recent theoretical developments include higher order perturbative QCD calculations \cite{Braun:2020yib} and a novel Monte Carlo event generator \cite{Aschenauer:2022aeb} for Deeply Virutal Compton Scattering (DVCS) and other exclusive processes. These new theoretical tools will soon be tested against existing  and future experimental data to extract the quark and gluon GPDs. 

However, even with  steady progress in theory and the promised capabilities of the EICs, the determination of the GPD $E$ still appears to be quite challenging, especially the gluon GPD $E_g$   \cite{Aschenauer:2013hhw}. The $E$-type GPDs  are associated with processes with  nucleon helicity-flip. It is commonly believed that such  processes are suppressed at high energy where the gluon degrees of freedom become important. At the moment, the distribution $E_g(x,\xi, t)$ ($x$ is the Bjorken variable, $\xi$ is the skewness parameter, $t$ is momentum transfer squared) is virtually unknown, and there is very little theory guidance on modeling it (see, e.g. \cite{Goloskokov:2008ib}). 

As a step forward to ameliorate this situation, in this paper we study the  small-$x$ behavior of the distribution $E_g(x)$ at vanishing momentum transfer $\xi=t=0$ from perturbative QCD point of view. For the nucleon helicity-nonflip GPD $H_g(x)$ at $\xi=0$, which is the same as the unpolarized gluon PDF, the Balitsky-Fadin-Kuraev-Lipatov (BFKL) resummation  technique \cite{Kuraev:1977fs,Balitsky:1978ic} is well established and widely used in phenomenology. The inclusion of the gluon saturation effects has also been extensively studied \cite{Gelis:2010nm}.  However, the corresponding framework for $E_g(x)$ does not seem to exist in the literature, even without gluon saturation.  We shall derive an evolution equation for  $xE_g(x)$ which resums the leading small-$x$ logarithms to all orders and determine the exponent $\delta$ of the Regge behavior $xE_g(x) \sim x^{-\delta}$.
%$(\alpha_s \ln 1/x)^n$ 
The impact of gluon saturation is also investigated.   Our result will be directly useful for phenomenology because forward gluon GPDs, or parton distribution functions (PDFs) in general, are often parameterized in the form $xf(x)\sim x^{-\delta}(1-x)^\beta$. Full GPDs $f(x,\xi,t)$ can then be modeled by common procedures \cite{Radyushkin:2000uy}. 

The small-$x$ behavior of $E_{g}(x)$ is also  important  for nucleon spin sum rules. In the Ji sum rule \cite{Ji:1996nm}, $E_g$ is related to the total angular momentum of gluons $J_g$ as 
 \begin{eqnarray}
 J_g =\frac{1}{2}\int_{0}^{1} \! d x \ x \left [H_g(x,\xi)+E_g(x,\xi)  \right ].\label{quark}
\end{eqnarray}
Depending on the value of $\delta$, there may be a significant contribution from $E_{g}$ to the integral. Similarly, in the  Jaffe-Manohar sum rule \cite{Jaffe:1989jz}, the canonical  gluon orbital angular momentum (OAM)  distribution ${\cal L}_{g}(x)$ is related to $E_{g}(x)$ as \cite{Hatta:2012cs}
\begin{eqnarray}
 {\cal L}_g(x) = x\int_x^1 \frac{dx'}{x'} (H_g(x')+E_g(x')) 
 \nonumber \\ 
 -\! 2x\! \int_x^1\! \frac{dx'}{x'^2} \Delta G(x')+\cdots, 
 \label{jm}
\end{eqnarray}
where $\Delta G(x)$ is the gluon helicity PDF and the genuine twist-three terms are omitted. Again, an enhancement of  $xE_{g}(x)\sim x^{-\delta}$ at small-$x$ induces a term ${\cal L}_{g}(x)\sim x^{-\delta}$ that can interfere with the contributions  from $H_g(x)$ and $\Delta G(x)$, and affect the extraction of ${\cal L}_g$ from experimental observables   \cite{Ji:2016jgn,Hatta:2016aoc,Bhattacharya:2017bvs,Bhattacharya:2018lgm,Bhattacharya:2022vvo}. 

On a broader perspective, our work is in line with the recent surge of activity towards understanding the small-$x$ QCD evolution of spin-related parton distributions via single $(\alpha_s \ln 1/x)^n$ and double $(\alpha_s\ln^2 1/x)^n$ logarithmic resummations  \cite{Kovchegov:2015pbl,Boussarie:2019icw,Chirilli:2021lif,Altinoluk:2021lvu,Tarasov:2021yll,Cougoulic:2022gbk}. 
As we shall see, the evolution of $E_g(x)$ is single-logarithmic despite its clear connection to the nucleon's spin degrees of freedom. In this sense it is similar to the evolution of the gluon Sivers function, or equivalently, the spin-dependent Odderon \cite{Zhou:2013gsa,Boer:2015pni,Dong:2018wsp,Yao:2018vcg,Boussarie:2019vmk,Hagiwara:2020mqb,Kovchegov:2021iyc,Boer:2022njw}. However, their small-$x$ behaviors turn out to be very different. \\

{\it Small-$x$ evolution equation for $E_g(x)$.} The standard way to determine the small-$x$ behavior of the unpolarized gluon PDF $xG(x) =xH_g(x,0,0)$ is to introduce the transverse momentum dependent (TMD) gluon distribution
$xG(x) = \int d^2k_\perp F(x,k_\perp)$ 
and solve the BFKL equation for $F(x,k_\perp)$. For the gluon GPD $E_g(x,0,0)$, the situation is more complicated  because it can only be defined through a nonforward matrix element $\langle p'|...|p\rangle$. Instead of a TMD, one has to consider the generalized TMD (GTMD) which depends on both $k_\perp$ and the momentum transfer $\Delta_\perp = p'_\perp-p_\perp$. As observed in \cite{Hatta:2016dxp}, the gluon GTMD at small-$x$ is proportional to the matrix element of the dipole S-matrix 
\begin{eqnarray}
\frac{1}{N_c} {\rm Tr} U(b_\perp+r_\perp/2) U^\dagger(b_\perp-r_\perp/2) ,
\end{eqnarray}
where $U(x_\perp)={\rm P}\exp\left(ig\int dz^-A^+(z^-,x_\perp)\right)
$ is a lightlike Wilson line in the fundamental representation. The dipole size $r_\perp$ is conjugate to $k_\perp$ and the impact parameter $b_\perp$ is conjugate to  $\Delta_\perp$. Importantly, the dipole S-matrix satisfies the small-$x$ evolution equation at the operator level \cite{Balitsky:1995ub}. Thus the equation remains valid after taking its matrix element between high energy proton states with definite  polarization.  In order to study the GPD $E_g$, it is natural to assume  transverse polarization and define
\begin{eqnarray}
  &&  \!\!\!\!\!  \!\!\!\!\! 2p^+ 2\pi \delta(p^+-p^{'+}) N(r_\perp,\Delta_\perp,S_\perp)\nonumber \\ &&
  \!\!\!\!\!  \!\!\!\!\!\equiv \langle p', S_\perp |1-\frac{1}{N_c}
  {\rm Tr} U(r_\perp/2) U^\dagger(-r_\perp/2) | p,S_\perp \rangle , \label{dipole}
  \end{eqnarray}
 where the polarization vector is normalized by the proton mass $|S_\perp|=M$. The above matrix element  
 can be  parameterized as  
 \cite{Boussarie:2019vmk,Hagiwara:2020mqb}.
  \begin{eqnarray}
  && \!\!\!\!\!
  \int \frac{d^2 r_\perp}{(2\pi)^2}
  e^{-ik_\perp \cdot r_\perp}\!  N(r_\perp,\Delta_\perp,S_\perp) 
  \approx \!(2\pi)^2\delta^{(2)}\!(\Delta_\perp)\delta^{(2)}\!(k_\perp) \! \nonumber \\ 
  &&\  - \frac{\pi g^2}{2N_c  k_\perp^2} \biggl[ 
   f_{1,1}
  -i
 \frac{ k_\perp \times S_\perp}{M^2}
\left(\frac{k_\perp \cdot \Delta_\perp}{M^2} f_{1,2}+ig_{1,2}\right) \nonumber \\&&
 \qquad  -i\frac{\Delta_\perp \times S_\perp}{2M^2}
  \left( 2f_{1,3}-f_{1,1}\right)\biggr],
\label{ret}
  \end{eqnarray}
  where  $  k_\perp \times S_\perp=\epsilon^{ij}k_{\perp}^i S^j_{\perp}$ and we have expanded the nucleon spinors to linear order in $\Delta_\perp$.  
  $f_{1,n}$ ($n=1,2,3$) are the real parts of the leading-twist gluon GTMDs at  $\xi\propto p^+-p'^+=0$. (See the delta function  in (\ref{dipole}). This is essentially the eikonal approximation.)  The imaginary parts have been omitted except for $g_{1,2}$ which is needed for a later discussion.  
  $f_{1,n}(k_\perp,\Delta_\perp=0)$ are related to the gluon GPDs through~\cite{Meissner:2009ww},
\begin{eqnarray}
xH_g(x)&=&\int d^2 k_\perp  f_{1,1}(k_\perp),
\\
xE_g(x)&=&\int d^2 k_\perp \bigl( -f_{1,1}(k_\perp)+ J(k_\perp) \bigr), \label{gpde}
\end{eqnarray}
where 
 \begin{eqnarray}
  J(k_\perp)\equiv 2f_{1,3}(k_\perp) + \frac{k_\perp^2}{M^2}f_{1,2}(k_\perp),
  \end{eqnarray}
is the TMD associated with the `angular momentum density'  $x(H_g(x)+E_g(x))$. Note that the above `tree-level' relations between the $k_\perp$-dependent and $k_\perp$-integrated distributions are modified by radiative corrections in the rigorous formulation of TMD  factorization~\cite{Collins:1984kg}.  However, this does not affect the determination of the small-$x$ asymptotic behavior in the leading logarithmic approximation, cf.,  \cite{Xiao:2017yya,Zhou:2018lfq}. (We  focus only on the small-$x$ logarithms and neglect the   rapidity (Sudakov) logarithms in TMDs. The latter can be included in future studies.)  

The nonlinear small-$x$ evolution equation for the dipole amplitude in momentum space is most conveniently expressed by a different Fourier transform from (\ref{ret}) 
\begin{eqnarray}
\!\! {\cal N}(k_\perp, \Delta_\perp)=
 \frac{1}{(2\pi)^2} \int d^2 r_\perp  e^{ik_\perp \cdot r_\perp } \frac{ { N}(r_\perp, \Delta_\perp)}{r_\perp^2}.
 \label{altf}
\end{eqnarray}
In terms of ${\cal N}$, the equation  reads\footnote{The equation is slightly different from  the one studied in \cite{Marquet:2005zf}. $\Delta_\perp$ here is Fourier conjugate to the impact parameter $b_\perp=(x_\perp+y_\perp)/2$ in the dipole amplitude $U(x_\perp)U^\dagger(y_\perp)$, whereas it is conjugate to $y_\perp$ in \cite{Marquet:2005zf}. }
\begin{eqnarray}
&& \!\!\!\!\!\!
\partial_Y {\cal N}(k_\perp, \Delta_\perp)=\frac{\bar \alpha_s}{\pi}\int \frac{d^2 k_\perp'}{(k_\perp-k_\perp')^2}  
\Biggl\{ {\cal N}(k_\perp', \Delta_\perp) 
 \nonumber \\ &&\left .\
   -\frac{1}{4}\left [ \frac{(\frac{\Delta_\perp}{2}+k_\perp)^2}{(\frac{\Delta_\perp}{2}+k_\perp')^2}+\frac{(\frac{\Delta_\perp}{2}-k_\perp)^2}{(\frac{\Delta_\perp}{2}-k_\perp')^2}\right ]{\cal N}(k_\perp, \Delta_\perp)  \right \} \nonumber \\ && -\frac{\bar{\alpha}_s}{2\pi} \int d^2\Delta'_\perp {\cal N}(k_\perp + \frac{\Delta'_\perp}{2},\Delta_\perp-\Delta'_\perp)\nonumber \\&&  \ \ \ \ \ \times {\cal N}(k_\perp + \frac{\Delta'_\perp-\Delta_\perp}{2},\Delta'_\perp),
 \label{bk}
  \end{eqnarray}
where  $\bar \alpha_s=\frac{\alpha_sN_c}{\pi}$ and $Y=\ln \frac{x_0}{x}$ is the rapidity. ($x_0$ is an arbitrary starting point of the evolution. Typically, $x_0\sim 0.01$.)  
%Note that in (\ref{altf}) we introduced a different Fourier transform than in (\ref{ret}). The sole reason for this  choice is  to simplify the nonlinear term in momentum space. 
 Similarly to (\ref{ret}), we parameterize ${\cal N}$ as 
\begin{eqnarray}
 && \!\!\!\!\!\!\!\!\!\!\!\!\!\!\!\!\!\!\!\!
 {\cal N}(k_\perp,\Delta_\perp,S_\perp)
  \approx    \frac{\pi g^2 }{2N_c  } \Biggl\{  
 {\cal F}_{1,1}(k_\perp,\Delta_\perp)\nonumber \\ &&\!\!\!\!\!\!\!\!\!\! -ik_\perp \times S_\perp \frac{k_\perp \cdot \Delta_\perp}{M^4}{\cal F}_{1,2}(k_\perp,\Delta_\perp) \nonumber \\ && \!\!\!\!\!\!\!\!\!\!-i\frac{\Delta_\perp \times S_\perp}{2M^2}\left[ 2{\cal F}_{1,3}(k_\perp,\Delta_\perp)-{\cal F}_{1,1}(k_\perp,\Delta_\perp)\right] \Biggr\},
 \label{newpara}
\end{eqnarray}
where the trivial (no-scattering) term has been omitted. The relation between ${\cal F}_{1,n}$ and $f_{i,n}$ can be easily worked out. In particular, we find 
\begin{eqnarray}
&& f_{1,1}(k_\perp)=k_\perp^2\frac{\partial^2}{\partial k_\perp^\alpha \partial k_\perp^\alpha}{\cal F}_{1,1}(k_\perp)  ,\nonumber\\
&& J(k_\perp) 
=k_\perp^2 \frac{\partial^2}{\partial k_\perp^\alpha \partial k_\perp^\alpha} \left(2{\cal F}_{1,3}(k_\perp) + \frac{k_\perp^2}{M^2} {\cal F}_{1,2}(k_\perp)\right)\nonumber \\ &&\qquad\quad\equiv k_\perp^2 \frac{\partial^2}{\partial k_\perp^\alpha \partial k_\perp^\alpha}{\cal J}(k_\perp).
\end{eqnarray}

We now substitute (\ref{newpara}) into (\ref{bk}) and separate the equation into the spin-independent and spin-dependent parts. The former is the standard Balitsky-Kovchegov (BK) equation with impact parameter  \cite{Balitsky:1995ub,Kovchegov:1999yj}.  Assuming a Gaussian profile ${\cal F}_{1,1}(k_\perp,\Delta_\perp)=\overline {\cal F}_{1,1}(k_\perp) {\cal A}_\perp e^{-\frac{\Delta_\perp^2 {\cal A}_\perp} {4\pi}}$ where ${\cal A}_\perp$ is the transverse area of the target, 
 we obtain   
%If we neglect the dependence on impact parameter and take the forward limit, we may write ${\cal  F}_{1,1}(k_\perp,\Delta_\perp)= \overline{{\cal F}}_{1,1}(k) (2\pi)^{2}\delta^{(2)}(\Delta_\perp )\to \overline{{\cal F}}_{1,1}(k_\perp) {\cal A}_\perp$ where ${\cal A}_\perp$ is the transverse area of the proton. The equation for $\overline{\cal F}_{1,1}$ reads
\begin{eqnarray}
  \partial_Y  \overline{\cal F}_{1,1}(k_\perp)&=&\frac{\bar \alpha_s}{\pi}\int \frac{d^2 k_\perp'}{(k_\perp-k_\perp')^2}  \biggl\{ \overline{\cal F}_{1,1}(k_\perp')     \label{bk1} \\
  &&  -\frac{1}{2} \frac{k_\perp^2}{k_\perp'^2}\overline{\cal F}_{1,1}(k_\perp) \biggr\}  - 4\pi^2\alpha_s^2  \left [\overline{\cal F}_{1,1}(k_\perp)\right  ]^2, \nonumber
\end{eqnarray}   
where we have taken the large area limit $ {\cal A}_\perp e^{-\frac{\Delta_\perp^2 {\cal A}_\perp} {4\pi}}\approx (2\pi)^2\delta^{(2)}(\Delta_\perp)$.  

The spin-dependent part is proportional to both  $S_\perp$ and $\Delta_\perp$.    Since $\Delta_\perp^i$ ($i=1,2$) are arbitrary,  we can remove this factor and recast the equation in  component form 
\begin{widetext}
\begin{eqnarray}
&&\!\!\!\!\!\!\! \!\!\!\! \!\!\!\!\!\partial_Y \left(k_\perp \times S_\perp \frac{k^i_\perp}{M^2}{\cal F}_{1,2}(k_\perp) + \epsilon^{ij}S^j_\perp  ({\cal F}_{1,3}(k_\perp)-\frac{1}{2}{\cal F}_{1,1}(k_\perp))  \right)
= \frac{\bar{\alpha}_s}{\pi} \int \frac{d^2k'_\perp}{(k_\perp-k'_\perp)^2} \Biggl[ k'_\perp \times S_\perp \frac{k'^i_\perp}{M^2}{\cal F}_{1,2}(k'_\perp)  \nonumber \\ 
&&+ \frac{\epsilon^{ij}S_\perp^j}{2} \bigl(2{\cal F}_{1,3}(k'_\perp)-{\cal F}_{1,1}(k'_\perp)\bigr)   - \frac{k_\perp^2}{2k'^2_\perp} \left( k_\perp \times S_\perp \frac{k_\perp^i}{M^2} {\cal F}_{1,2}(k_\perp) +\frac{\epsilon^{ij}S^j_\perp}{2} \bigl(2{\cal F}_{1,3}(k_\perp)-{\cal F}_{1,1}(k_\perp)\bigr)\right) \Biggr] \nonumber \\ 
&& -4\pi^2\alpha_s^2 \left(k_\perp\times S_\perp\frac{k_\perp^i}{M^2}{\cal F}_{1,2}(k_\perp) + \frac{\epsilon^{ij}S_\perp^j}{2} \bigl(2{\cal F}_{1,3}(k_\perp)-{\cal F}_{1,1}(k_\perp) \bigr) \right)\overline{\cal F}_{1,1}(k_\perp) , \label{mast}
\end{eqnarray} 
\end{widetext}
where we assumed that ${\cal F}_{1,n}$ $(n=1,2,3)$ have the same Gaussian width and wrote ${\cal F}_{1,n}(k_\perp,\Delta_\perp)= {\cal F}_{1,n}(k_\perp)e^{-\frac{\Delta_\perp^2{\cal A}_\perp}{4\pi}}$.
%The overall prefactor of the nonlinear terms (the last line of (\ref{mast})) is to some extent model-dependent\footnote{\new{For instance, if we assume that the $\Delta_\perp$ dependence of  the GTMDs  take a Gaussian form ${\cal F}_{1,1} \propto e^{-\Delta_\perp^2/A}$ and ${\cal F}_{1,3}\propto  e^{-\Delta_\perp^2/B}$, the resulting coefficient for the nonlinear term is $\frac{2B}{A+B}4\pi^2 \alpha_s^2$ after carrying out $\Delta_\perp'$ integration. }} but with different widths. Here we have assumed that ${\cal F}_{1,n}(k_\perp,\Delta_\perp)$ with $n=1,2,3$ are sharply peaked around  $\Delta_\perp=0$, and that all have the same width in $\Delta_\perp$.  
We can decompose (\ref{mast}) into independent basis vectors and obtain the following two evolution equations 
\begin{eqnarray}
\!\!\!\!\!\!\!\!\!\!\!
\partial_Y {\cal F}_{1,2}(k_\perp) &=& \frac{\bar{\alpha}_s}{\pi} \int   \frac{d^2k'_\perp}{(k_\perp-k'_\perp)^2} \left[ -\frac{k_\perp^2}{2k'^2_\perp} {\cal F}_{1,2}(k_\perp)\right .\ \nonumber \\ &&\!\!\!\!\!\!\!\!\!\!\!\left .\ + \frac{2(k_\perp \cdot k'_\perp)^2-k_\perp^2 k'^2_\perp}{(k_\perp^2)^2} {\cal F}_{1,2}(k'_\perp)\right] \nonumber \\ 
&&\!\!\!\!\!\!\!\!\!\!\! -4\pi^2\alpha_s^2 \overline{\cal F}_{1,1}(k_\perp){\cal F}_{1,2}(k_\perp) , \label{f12}
\end{eqnarray}
and 
\begin{eqnarray}
\partial_Y {\cal F}_{1,3}(k_\perp)  &=& \frac{\bar{\alpha}_s}{\pi} \int \frac{d^2k'_\perp}{(k_\perp-k'_\perp)^2} \Biggl[   - \frac{k_\perp^2}{2k'^2_\perp} {\cal F}_{1,3}(k_\perp)\nonumber \\&&\!\!\!\!\!\!\!\!\!\!\! +\frac{k_\perp^2 k'^2_\perp -(k_\perp\cdot k'_\perp)^2}{k_\perp^2} \frac{{\cal F}_{1,2}(k'_\perp)}{M^2} + {\cal F}_{1,3}(k_\perp')\Biggr] \nonumber \\ 
&& \!\!\!\!\!\!\!\!\!\!\!-4\pi^2\alpha_s^2 \overline{\cal F}_{1,1}(k_\perp){\cal F}_{1,3}(k_\perp). \label{f13}
\end{eqnarray}
In the `dilute' regime where the effect of gluon saturation is negligible, the last term in (\ref{f12}) and (\ref{f13}) can be omitted.  ${\cal F}_{1,2}$ then satisfies a closed evolution equation whereas ${\cal F}_{1,3}$ does not.  
Combining these two equations, we arrive at the main result of this paper 
\begin{eqnarray}
\partial_Y {\cal J}(k_\perp)  &=&   \frac{\bar{\alpha}_s}{\pi} \!\int \frac{d^2k'_\perp}{(k_\perp-k'_\perp)^2} \Biggl[ {\cal J}(k_\perp')  - \frac{k_\perp^2}{2k'^2_\perp} {\cal J}(k_\perp) \Biggr] \nonumber \\
&& -4\pi^2\alpha_s^2 \overline{\cal F}_{1,1}(k_\perp){\cal J}(k_\perp). \label{jj}
\end{eqnarray} 
If we define ${\cal E}\equiv -{\cal F}_{1,1}+{\cal J}$ such that $k_\perp^2 \frac{\partial^2}{\partial k_\perp^\alpha \partial k_\perp^\alpha} {\cal E} = -f_{1,1}+J$ is the TMD associated with $xE_g(x)$ (see (\ref{gpde})),  we similarly obtain 
\begin{eqnarray}
\partial_Y {\cal E}(k_\perp)  &=&   \frac{\bar{\alpha}_s}{\pi} \int \frac{d^2k'_\perp}{(k_\perp-k'_\perp)^2} \Biggl[ {\cal E}(k_\perp')  - \frac{k_\perp^2}{2k'^2_\perp} {\cal E}(k_\perp) \Biggr] \nonumber \\
&& -4\pi^2\alpha_s^2 \overline{\cal F}_{1,1}(k_\perp){\cal E}(k_\perp). \label{e}
\end{eqnarray} 
The linear part of (\ref{f13}) and (\ref{e})  is nothing but the BFKL equation. We thus conclude that   ${\cal E}(k_\perp)$, hence also $xE_g(x)$ behave like the BKFL Pomeron \cite{Kuraev:1977fs,Balitsky:1978ic} in the same way as the unpolarized gluon PDF 
\begin{eqnarray}
xE_g(x)\sim  xG(x) \propto \left(\frac{1}{x}\right)^{\bar{\alpha}_s4\ln 2}. \label{bfk}
\end{eqnarray}
Gluon saturation as represented by the `nonlinear' term  suppresses the rapid BFKL growth. Note however that it has the form ${\cal F}_{1,1}{\cal E}$ instead of ${\cal E}{\cal E}$ as one would naively expect from (\ref{bk1}). Consequently, in the $x\to 0$ limit the ratio
\begin{eqnarray}
R\equiv \frac{{\cal E}(x,k_\perp)}{ {\cal F}_{1,1}(x,k_\perp)},  \label{esat}
\end{eqnarray}
approaches a constant independent of $x$ and $k_\perp$.

It is worthwhile to compare (\ref{bfk}) and  (\ref{esat})  with the behavior of the gluon Sivers function, or the `spin-dependent Odderon'  $xg_{1,2}(x,k_\perp)$, see (\ref{ret}). Although both ${\cal E}$ and $xg_{1,2}$ are associated with transverse polarization and satisfy the same BFKL equation in the dilute regime, their small-$x$ behaviors  are drastically different. This is because the $g_{1,2}$ term is odd in $k_\perp$, and selects the subleading (Odderon) solution of the BFKL equation $xg_{1,2}\sim (1/x)^0$ \cite{Bartels:1999yt}.  Moreover, in the saturation regime, $xg_{1,2}$ gets strongly suppressed   \cite{Lappi:2016gqe,Yao:2018vcg,Contreras:2020lrh}. 
\\
 
\begin{figure}[t]\centering
\includegraphics[width=0.4\textwidth]{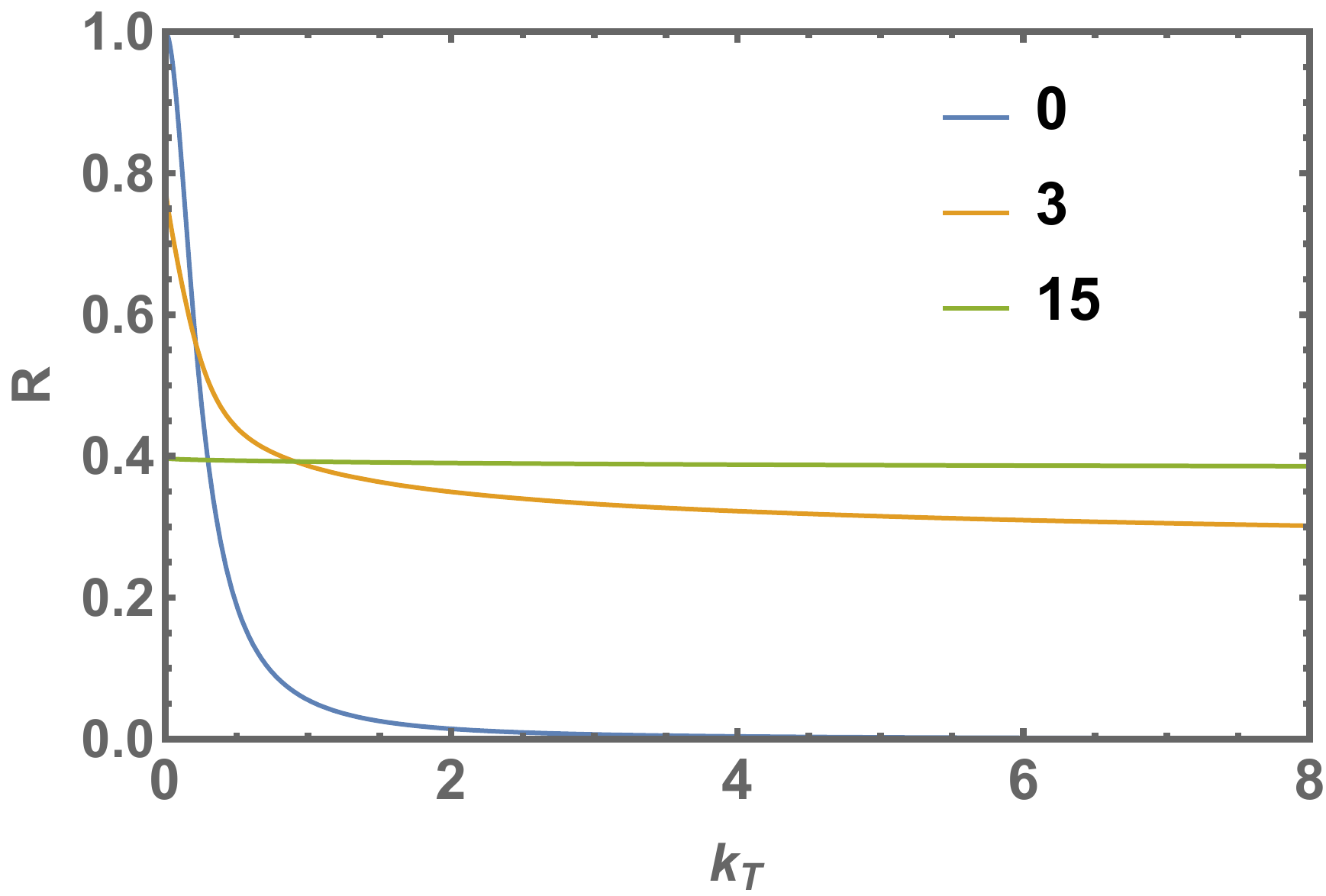}
\includegraphics[width=0.4\textwidth]{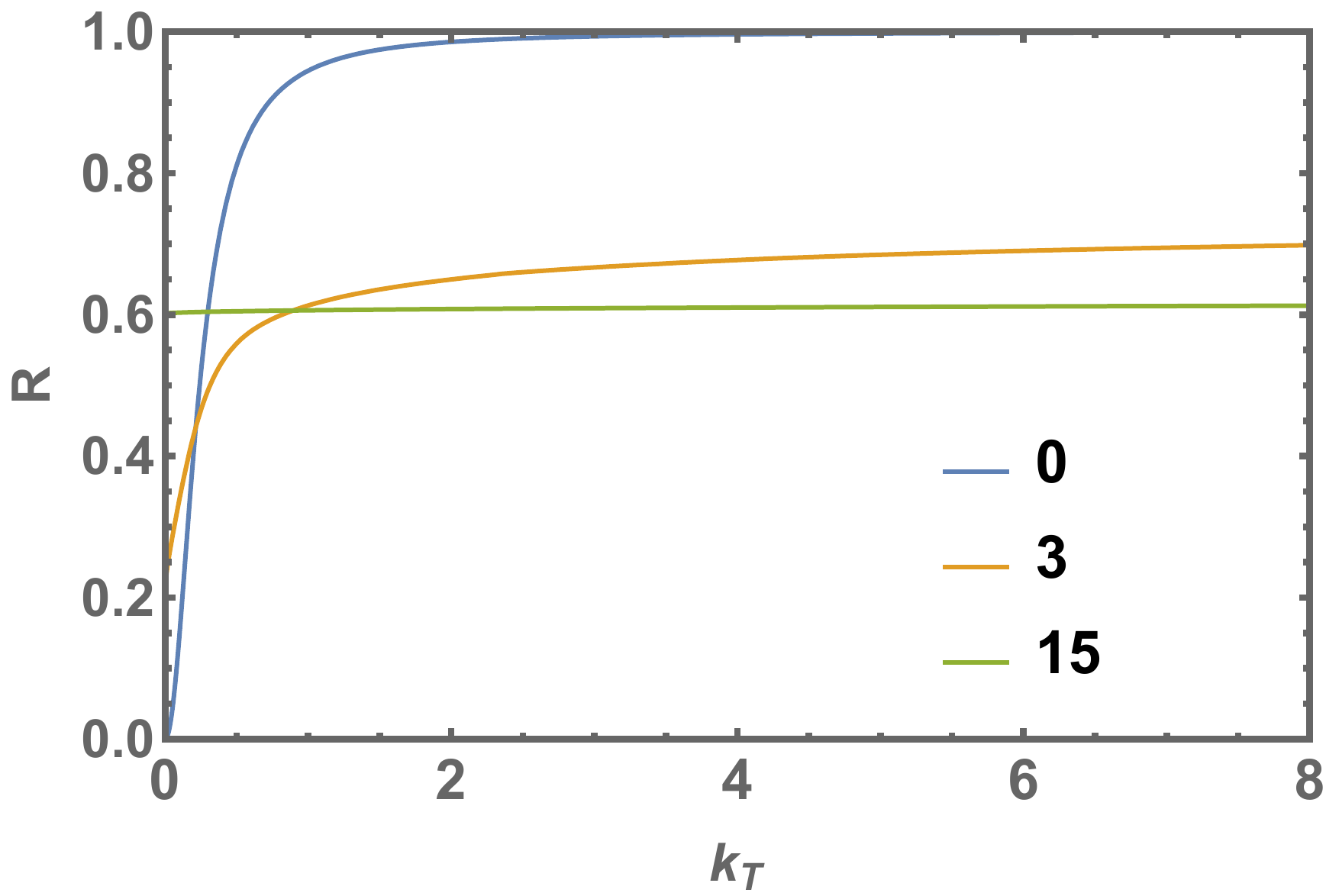}
\caption{ The ratio (\ref{esat}) plotted as a  function of $k_\perp$ (in GeV) at $Y=0, 3 ,15$
 for two different initial conditions for ${\cal E}(k_\perp)$, (\ref{model1}) (top panel) and (\ref{model2}) (bottom panel).
}
\label{fig:lhcvsrhic}
\end{figure}

{\it Numerical results.}
We numerically solve the coupled evolution equations (\ref{bk1}) and (\ref{e}) using the McLerran-Venugopalan model~\cite{McLerran:1993ni} as the initial condition for ${\cal F}_{1,1}(k_\perp)$ at $Y=0$ 
\begin{eqnarray}
{\cal F}_{1,1}(Y=0,k_\perp)&=& \frac{N_c{\cal A}_\perp}{2\pi^2 \alpha_s } \int \frac{d^2 r_\perp}{(2\pi)^2 r_\perp^2} e^{-ik_\perp \cdot r_\perp} \\&& \!\!\!\!\!\!\!\!\!\!\!\!\!\!\! \!\!\!\!\!\!\!\ \times\left \{1-\exp \left [-\frac{r_\perp^2 Q_{s0}^2}{4}\ln\left( \frac{1}{r_\perp \Lambda_{\text{mv}}}+e \right ) \right ] \right \}, \nonumber
\end{eqnarray}
where   $Q_{0s}^2= \ 0.2 {\text{ GeV}}^2$ is the initial saturation scale and 
 $\Lambda_\text{mv}=0.241$ GeV is an infrared cutoff. As for ${\cal E}$, we employ two models  
\begin{eqnarray}
\!\!\!\!\!\!\!\!\!\!\!\!\!\!\!\!\!\!&&
 {\cal E}(Y=0,k_\perp)=\frac{\Lambda_{\text{mv}}^2}{k_\perp^2+\Lambda_{\text{mv}}^2 }  {\cal F}_{1,1}(Y=0,k_\perp) , \label{model1} \\
\!\!\!\!\!\!\!\!\!\!\!\!\!\!\!\!\!\!&&
{\cal E}(Y=0,k_\perp)=\frac{k_\perp^2}{k_\perp^2+\Lambda_{\text{mv}}^2}   {\cal F}_{1,1}(Y=0,k_\perp) . \label{model2}
\end{eqnarray}
We use a fixed coupling constant $\alpha_s=0.2$. 
The obtained ratio (\ref{esat}) is plotted in Fig.~\ref{fig:lhcvsrhic} as a function of $k_\perp$ at different values of $Y$.  As expected,   at large enough rapidities, ${\cal E}(k_\perp)$ and  ${\cal F}_{1,1}(k_\perp)$ have the same $k_\perp$-distribution  so that their ratio is frozen. The asymptotic value of $R$ depends on the initial condition, but it does not depend on $\alpha_s$. The latter only affects the speed at which the asymptotic value is reached. This means that in the deeply saturated regime, not only  ${\cal F}_{1,1}$, but also  ${\cal E}$ attain the `high density fixed point' 
\begin{eqnarray}
{\cal E} \propto {\cal F}_{1,1} \sim  {\cal O}\left(\frac{1}{\alpha_s}\right), \label{cgc}
\end{eqnarray}
characteristic of gluon saturation \cite{Gelis:2010nm}.  \\

{\it Conclusion.}
In this work, we have derived the small-$x$ evolution equation (\ref{e}) for the TMD ${\cal E}(x,k_\perp)$ of the gluon GPD $E_g(x)$. An immediate consequence is that $E_g(x)$ grows rapidly at small-$x$ in exactly the same way as the BFKL Pomeron. This growth is eventually damped by gluon saturation, and the ratio $E_g(x)/H_g(x)$ goes to a constant in the limit $x\to 0$. It is remarkable that the nucleon helicity-flip distribution  behaves exactly like the  helicity-nonflip one.  The former is often assumed to be negligible at high energy \cite{Goloskokov:2008ib}.

Our result sends a clear and lasting message for the  modeling of  $E_g(x,\xi)$. The exponent $\delta$ in $xE_g(x)\sim x^{-\delta}$ must be the same as that for $xH_g(x)=xG(x)$. While such a behavior was  occasionally postulated in the literature based on pure assumption, a systematic derivation of this result has been achieved for the first time. A common choice in the small-$x$ literature is $\delta\sim 0.3$, though  smaller values may be preferred  at low renormalization scales in view of the well-known transition between the hard and soft Pomerons \cite{H1:2001ert}.

The strong rise  of $xE_g(x)$  can induce large  transverse single spin asymmetries in various exclusive processes. As an  example, for exclusive $J/\psi$ production in ultraperipheral  collisions (UPC),  a sizable asymmetry was predicted from a conservative choice  $\delta=0.1$  \cite{Koempel:2011rc,Lansberg:2018fsy}. With larger values of $\delta$, the asymmetry gets  even larger and could be measured at RHIC and the LHC.  
Our result also suggests that there may be a significant contribution to nucleon spin sum rules (\ref{quark}) and (\ref{jm}) from $E_g(x)$ in the small-$x$ region. In particular, this will  affect the cross section of observables related to the gluon canonical OAM ${\cal L}_g(x)$  \cite{Bhattacharya:2022vvo}. It remains to be seen whether similar conclusions can be drawn for the quark GPD $E_q(x)$ and the quark OAMs, or whether the scaling (\ref{cgc}) holds after including the next-to-leading logarithmic corrections. These issues are particularly important for phenomenology at RHIC and the future EICs where the collision energies are not asymptotically high.

\section*{Acknowledgments} 
We thank Feng Yuan for bringing our attention to this topic,  and Lech Szymanowski and Bowen Xiao for discussions. Y. H. is supported by the U.~S. Department of Energy, Office of Science, Office of Nuclear Physics under contract number DE- SC0012704,  and also by  Laboratory Directed Research and Development (LDRD) funds from Brookhaven Science Associates.
The work of J.~Z. is  supported by the National Natural Science Foundations of China under Grant No. 1217511.

\bibliography{ref.bib}

\begin{thebibliography}{48}
\expandafter\ifx\csname natexlab\endcsname\relax\def\natexlab#1{#1}\fi
\expandafter\ifx\csname bibnamefont\endcsname\relax
  \def\bibnamefont#1{#1}\fi
\expandafter\ifx\csname bibfnamefont\endcsname\relax
  \def\bibfnamefont#1{#1}\fi
\expandafter\ifx\csname citenamefont\endcsname\relax
  \def\citenamefont#1{#1}\fi
\expandafter\ifx\csname url\endcsname\relax
  \def\url#1{\texttt{#1}}\fi
\expandafter\ifx\csname urlprefix\endcsname\relax\def\urlprefix{URL }\fi
\providecommand{\bibinfo}[2]{#2}
\providecommand{\eprint}[2][]{\url{#2}}

\bibitem[{\citenamefont{Abdul~Khalek et~al.}(2021)}]{AbdulKhalek:2021gbh}
\bibinfo{author}{\bibfnamefont{R.}~\bibnamefont{Abdul~Khalek}}
  \bibnamefont{et~al.} (\bibinfo{year}{2021}), \eprint{2103.05419}.

\bibitem[{\citenamefont{Anderle et~al.}(2021)}]{Anderle:2021wcy}
\bibinfo{author}{\bibfnamefont{D.~P.} \bibnamefont{Anderle}}
  \bibnamefont{et~al.}, \bibinfo{journal}{Front. Phys. (Beijing)}
  \textbf{\bibinfo{volume}{16}}, \bibinfo{pages}{64701} (\bibinfo{year}{2021}),
  \eprint{2102.09222}.

\bibitem[{\citenamefont{Diehl}(2003)}]{Diehl:2003ny}
\bibinfo{author}{\bibfnamefont{M.}~\bibnamefont{Diehl}},
  \bibinfo{journal}{Phys. Rept.} \textbf{\bibinfo{volume}{388}},
  \bibinfo{pages}{41} (\bibinfo{year}{2003}), \eprint{hep-ph/0307382}.

\bibitem[{\citenamefont{Braun et~al.}(2020)\citenamefont{Braun, Manashov, Moch,
  and Schoenleber}}]{Braun:2020yib}
\bibinfo{author}{\bibfnamefont{V.~M.} \bibnamefont{Braun}},
  \bibinfo{author}{\bibfnamefont{A.~N.} \bibnamefont{Manashov}},
  \bibinfo{author}{\bibfnamefont{S.}~\bibnamefont{Moch}}, \bibnamefont{and}
  \bibinfo{author}{\bibfnamefont{J.}~\bibnamefont{Schoenleber}},
  \bibinfo{journal}{JHEP} \textbf{\bibinfo{volume}{09}}, \bibinfo{pages}{117}
  (\bibinfo{year}{2020}), \eprint{2007.06348}.

\bibitem[{\citenamefont{Aschenauer et~al.}(2022)\citenamefont{Aschenauer,
  Batozskaya, Fazio, Gates, Moutarde, Sokhan, Spiesberger, Sznajder, and
  Tezgin}}]{Aschenauer:2022aeb}
\bibinfo{author}{\bibfnamefont{E.~C.} \bibnamefont{Aschenauer}},
  \bibinfo{author}{\bibfnamefont{V.}~\bibnamefont{Batozskaya}},
  \bibinfo{author}{\bibfnamefont{S.}~\bibnamefont{Fazio}},
  \bibinfo{author}{\bibfnamefont{K.}~\bibnamefont{Gates}},
  \bibinfo{author}{\bibfnamefont{H.}~\bibnamefont{Moutarde}},
  \bibinfo{author}{\bibfnamefont{D.}~\bibnamefont{Sokhan}},
  \bibinfo{author}{\bibfnamefont{H.}~\bibnamefont{Spiesberger}},
  \bibinfo{author}{\bibfnamefont{P.}~\bibnamefont{Sznajder}}, \bibnamefont{and}
  \bibinfo{author}{\bibfnamefont{K.}~\bibnamefont{Tezgin}}
  (\bibinfo{year}{2022}), \eprint{2205.01762}.

\bibitem[{\citenamefont{Aschenauer et~al.}(2013)\citenamefont{Aschenauer,
  Fazio, Kumericki, and Mueller}}]{Aschenauer:2013hhw}
\bibinfo{author}{\bibfnamefont{E.-C.} \bibnamefont{Aschenauer}},
  \bibinfo{author}{\bibfnamefont{S.}~\bibnamefont{Fazio}},
  \bibinfo{author}{\bibfnamefont{K.}~\bibnamefont{Kumericki}},
  \bibnamefont{and} \bibinfo{author}{\bibfnamefont{D.}~\bibnamefont{Mueller}},
  \bibinfo{journal}{JHEP} \textbf{\bibinfo{volume}{09}}, \bibinfo{pages}{093}
  (\bibinfo{year}{2013}), \eprint{1304.0077}.

\bibitem[{\citenamefont{Goloskokov and Kroll}(2009)}]{Goloskokov:2008ib}
\bibinfo{author}{\bibfnamefont{S.~V.} \bibnamefont{Goloskokov}}
  \bibnamefont{and} \bibinfo{author}{\bibfnamefont{P.}~\bibnamefont{Kroll}},
  \bibinfo{journal}{Eur. Phys. J. C} \textbf{\bibinfo{volume}{59}},
  \bibinfo{pages}{809} (\bibinfo{year}{2009}), \eprint{0809.4126}.

\bibitem[{\citenamefont{Kuraev et~al.}(1977)\citenamefont{Kuraev, Lipatov, and
  Fadin}}]{Kuraev:1977fs}
\bibinfo{author}{\bibfnamefont{E.~A.} \bibnamefont{Kuraev}},
  \bibinfo{author}{\bibfnamefont{L.~N.} \bibnamefont{Lipatov}},
  \bibnamefont{and} \bibinfo{author}{\bibfnamefont{V.~S.} \bibnamefont{Fadin}},
  \bibinfo{journal}{Sov. Phys. JETP} \textbf{\bibinfo{volume}{45}},
  \bibinfo{pages}{199} (\bibinfo{year}{1977}).

\bibitem[{\citenamefont{Balitsky and Lipatov}(1978)}]{Balitsky:1978ic}
\bibinfo{author}{\bibfnamefont{I.~I.} \bibnamefont{Balitsky}} \bibnamefont{and}
  \bibinfo{author}{\bibfnamefont{L.~N.} \bibnamefont{Lipatov}},
  \bibinfo{journal}{Sov. J. Nucl. Phys.} \textbf{\bibinfo{volume}{28}},
  \bibinfo{pages}{822} (\bibinfo{year}{1978}).

\bibitem[{\citenamefont{Gelis et~al.}(2010)\citenamefont{Gelis, Iancu,
  Jalilian-Marian, and Venugopalan}}]{Gelis:2010nm}
\bibinfo{author}{\bibfnamefont{F.}~\bibnamefont{Gelis}},
  \bibinfo{author}{\bibfnamefont{E.}~\bibnamefont{Iancu}},
  \bibinfo{author}{\bibfnamefont{J.}~\bibnamefont{Jalilian-Marian}},
  \bibnamefont{and}
  \bibinfo{author}{\bibfnamefont{R.}~\bibnamefont{Venugopalan}},
  \bibinfo{journal}{Ann. Rev. Nucl. Part. Sci.} \textbf{\bibinfo{volume}{60}},
  \bibinfo{pages}{463} (\bibinfo{year}{2010}), \eprint{1002.0333}.

\bibitem[{\citenamefont{Radyushkin}(2000)}]{Radyushkin:2000uy}
\bibinfo{author}{\bibfnamefont{A.~V.} \bibnamefont{Radyushkin}}
  (\bibinfo{year}{2000}), \eprint{hep-ph/0101225}.

\bibitem[{\citenamefont{Ji}(1997)}]{Ji:1996nm}
\bibinfo{author}{\bibfnamefont{X.-D.} \bibnamefont{Ji}},
  \bibinfo{journal}{Phys. Rev. D} \textbf{\bibinfo{volume}{55}},
  \bibinfo{pages}{7114} (\bibinfo{year}{1997}), \eprint{hep-ph/9609381}.

\bibitem[{\citenamefont{Jaffe and Manohar}(1990)}]{Jaffe:1989jz}
\bibinfo{author}{\bibfnamefont{R.~L.} \bibnamefont{Jaffe}} \bibnamefont{and}
  \bibinfo{author}{\bibfnamefont{A.}~\bibnamefont{Manohar}},
  \bibinfo{journal}{Nucl. Phys. B} \textbf{\bibinfo{volume}{337}},
  \bibinfo{pages}{509} (\bibinfo{year}{1990}).

\bibitem[{\citenamefont{Hatta and Yoshida}(2012)}]{Hatta:2012cs}
\bibinfo{author}{\bibfnamefont{Y.}~\bibnamefont{Hatta}} \bibnamefont{and}
  \bibinfo{author}{\bibfnamefont{S.}~\bibnamefont{Yoshida}},
  \bibinfo{journal}{JHEP} \textbf{\bibinfo{volume}{10}}, \bibinfo{pages}{080}
  (\bibinfo{year}{2012}), \eprint{1207.5332}.

\bibitem[{\citenamefont{Ji et~al.}(2017)\citenamefont{Ji, Yuan, and
  Zhao}}]{Ji:2016jgn}
\bibinfo{author}{\bibfnamefont{X.}~\bibnamefont{Ji}},
  \bibinfo{author}{\bibfnamefont{F.}~\bibnamefont{Yuan}}, \bibnamefont{and}
  \bibinfo{author}{\bibfnamefont{Y.}~\bibnamefont{Zhao}},
  \bibinfo{journal}{Phys. Rev. Lett.} \textbf{\bibinfo{volume}{118}},
  \bibinfo{pages}{192004} (\bibinfo{year}{2017}), \eprint{1612.02438}.

\bibitem[{\citenamefont{Hatta et~al.}(2017)\citenamefont{Hatta, Nakagawa, Yuan,
  Zhao, and Xiao}}]{Hatta:2016aoc}
\bibinfo{author}{\bibfnamefont{Y.}~\bibnamefont{Hatta}},
  \bibinfo{author}{\bibfnamefont{Y.}~\bibnamefont{Nakagawa}},
  \bibinfo{author}{\bibfnamefont{F.}~\bibnamefont{Yuan}},
  \bibinfo{author}{\bibfnamefont{Y.}~\bibnamefont{Zhao}}, \bibnamefont{and}
  \bibinfo{author}{\bibfnamefont{B.}~\bibnamefont{Xiao}},
  \bibinfo{journal}{Phys. Rev. D} \textbf{\bibinfo{volume}{95}},
  \bibinfo{pages}{114032} (\bibinfo{year}{2017}), \eprint{1612.02445}.

\bibitem[{\citenamefont{Bhattacharya et~al.}(2017)\citenamefont{Bhattacharya,
  Metz, and Zhou}}]{Bhattacharya:2017bvs}
\bibinfo{author}{\bibfnamefont{S.}~\bibnamefont{Bhattacharya}},
  \bibinfo{author}{\bibfnamefont{A.}~\bibnamefont{Metz}}, \bibnamefont{and}
  \bibinfo{author}{\bibfnamefont{J.}~\bibnamefont{Zhou}},
  \bibinfo{journal}{Phys. Lett. B} \textbf{\bibinfo{volume}{771}},
  \bibinfo{pages}{396} (\bibinfo{year}{2017}), \bibinfo{note}{[Erratum:
  Phys.Lett.B 810, 135866 (2020)]}, \eprint{1702.04387}.

\bibitem[{\citenamefont{Bhattacharya et~al.}(2018)\citenamefont{Bhattacharya,
  Metz, Ojha, Tsai, and Zhou}}]{Bhattacharya:2018lgm}
\bibinfo{author}{\bibfnamefont{S.}~\bibnamefont{Bhattacharya}},
  \bibinfo{author}{\bibfnamefont{A.}~\bibnamefont{Metz}},
  \bibinfo{author}{\bibfnamefont{V.~K.} \bibnamefont{Ojha}},
  \bibinfo{author}{\bibfnamefont{J.-Y.} \bibnamefont{Tsai}}, \bibnamefont{and}
  \bibinfo{author}{\bibfnamefont{J.}~\bibnamefont{Zhou}}
  (\bibinfo{year}{2018}), \eprint{1802.10550}.

\bibitem[{\citenamefont{Bhattacharya et~al.}(2022)\citenamefont{Bhattacharya,
  Boussarie, and Hatta}}]{Bhattacharya:2022vvo}
\bibinfo{author}{\bibfnamefont{S.}~\bibnamefont{Bhattacharya}},
  \bibinfo{author}{\bibfnamefont{R.}~\bibnamefont{Boussarie}},
  \bibnamefont{and} \bibinfo{author}{\bibfnamefont{Y.}~\bibnamefont{Hatta}},
  \bibinfo{journal}{Phys. Rev. Lett.} \textbf{\bibinfo{volume}{128}},
  \bibinfo{pages}{182002} (\bibinfo{year}{2022}), \eprint{2201.08709}.

\bibitem[{\citenamefont{Kovchegov et~al.}(2016)\citenamefont{Kovchegov,
  Pitonyak, and Sievert}}]{Kovchegov:2015pbl}
\bibinfo{author}{\bibfnamefont{Y.~V.} \bibnamefont{Kovchegov}},
  \bibinfo{author}{\bibfnamefont{D.}~\bibnamefont{Pitonyak}}, \bibnamefont{and}
  \bibinfo{author}{\bibfnamefont{M.~D.} \bibnamefont{Sievert}},
  \bibinfo{journal}{JHEP} \textbf{\bibinfo{volume}{01}}, \bibinfo{pages}{072}
  (\bibinfo{year}{2016}), \bibinfo{note}{[Erratum: JHEP 10, 148 (2016)]},
  \eprint{1511.06737}.

\bibitem[{\citenamefont{Boussarie et~al.}(2019)\citenamefont{Boussarie, Hatta,
  and Yuan}}]{Boussarie:2019icw}
\bibinfo{author}{\bibfnamefont{R.}~\bibnamefont{Boussarie}},
  \bibinfo{author}{\bibfnamefont{Y.}~\bibnamefont{Hatta}}, \bibnamefont{and}
  \bibinfo{author}{\bibfnamefont{F.}~\bibnamefont{Yuan}},
  \bibinfo{journal}{Phys. Lett. B} \textbf{\bibinfo{volume}{797}},
  \bibinfo{pages}{134817} (\bibinfo{year}{2019}), \eprint{1904.02693}.

\bibitem[{\citenamefont{Chirilli}(2021)}]{Chirilli:2021lif}
\bibinfo{author}{\bibfnamefont{G.~A.} \bibnamefont{Chirilli}},
  \bibinfo{journal}{JHEP} \textbf{\bibinfo{volume}{06}}, \bibinfo{pages}{096}
  (\bibinfo{year}{2021}), \eprint{2101.12744}.

\bibitem[{\citenamefont{Altinoluk and Beuf}(2022)}]{Altinoluk:2021lvu}
\bibinfo{author}{\bibfnamefont{T.}~\bibnamefont{Altinoluk}} \bibnamefont{and}
  \bibinfo{author}{\bibfnamefont{G.}~\bibnamefont{Beuf}},
  \bibinfo{journal}{Phys. Rev. D} \textbf{\bibinfo{volume}{105}},
  \bibinfo{pages}{074026} (\bibinfo{year}{2022}), \eprint{2109.01620}.

\bibitem[{\citenamefont{Tarasov and Venugopalan}(2022)}]{Tarasov:2021yll}
\bibinfo{author}{\bibfnamefont{A.}~\bibnamefont{Tarasov}} \bibnamefont{and}
  \bibinfo{author}{\bibfnamefont{R.}~\bibnamefont{Venugopalan}},
  \bibinfo{journal}{Phys. Rev. D} \textbf{\bibinfo{volume}{105}},
  \bibinfo{pages}{014020} (\bibinfo{year}{2022}), \eprint{2109.10370}.

\bibitem[{\citenamefont{Cougoulic et~al.}(2022)\citenamefont{Cougoulic,
  Kovchegov, Tarasov, and Tawabutr}}]{Cougoulic:2022gbk}
\bibinfo{author}{\bibfnamefont{F.}~\bibnamefont{Cougoulic}},
  \bibinfo{author}{\bibfnamefont{Y.~V.} \bibnamefont{Kovchegov}},
  \bibinfo{author}{\bibfnamefont{A.}~\bibnamefont{Tarasov}}, \bibnamefont{and}
  \bibinfo{author}{\bibfnamefont{Y.}~\bibnamefont{Tawabutr}}
  (\bibinfo{year}{2022}), \eprint{2204.11898}.

\bibitem[{\citenamefont{Zhou}(2014)}]{Zhou:2013gsa}
\bibinfo{author}{\bibfnamefont{J.}~\bibnamefont{Zhou}}, \bibinfo{journal}{Phys.
  Rev. D} \textbf{\bibinfo{volume}{89}}, \bibinfo{pages}{074050}
  (\bibinfo{year}{2014}), \eprint{1308.5912}.

\bibitem[{\citenamefont{Boer et~al.}(2016)\citenamefont{Boer, Echevarria,
  Mulders, and Zhou}}]{Boer:2015pni}
\bibinfo{author}{\bibfnamefont{D.}~\bibnamefont{Boer}},
  \bibinfo{author}{\bibfnamefont{M.~G.} \bibnamefont{Echevarria}},
  \bibinfo{author}{\bibfnamefont{P.}~\bibnamefont{Mulders}}, \bibnamefont{and}
  \bibinfo{author}{\bibfnamefont{J.}~\bibnamefont{Zhou}},
  \bibinfo{journal}{Phys. Rev. Lett.} \textbf{\bibinfo{volume}{116}},
  \bibinfo{pages}{122001} (\bibinfo{year}{2016}), \eprint{1511.03485}.

\bibitem[{\citenamefont{Dong et~al.}(2019)\citenamefont{Dong, Zheng, and
  Zhou}}]{Dong:2018wsp}
\bibinfo{author}{\bibfnamefont{H.}~\bibnamefont{Dong}},
  \bibinfo{author}{\bibfnamefont{D.-X.} \bibnamefont{Zheng}}, \bibnamefont{and}
  \bibinfo{author}{\bibfnamefont{J.}~\bibnamefont{Zhou}},
  \bibinfo{journal}{Phys. Lett. B} \textbf{\bibinfo{volume}{788}},
  \bibinfo{pages}{401} (\bibinfo{year}{2019}), \eprint{1805.09479}.

\bibitem[{\citenamefont{Yao et~al.}(2019)\citenamefont{Yao, Hagiwara, and
  Hatta}}]{Yao:2018vcg}
\bibinfo{author}{\bibfnamefont{X.}~\bibnamefont{Yao}},
  \bibinfo{author}{\bibfnamefont{Y.}~\bibnamefont{Hagiwara}}, \bibnamefont{and}
  \bibinfo{author}{\bibfnamefont{Y.}~\bibnamefont{Hatta}},
  \bibinfo{journal}{Phys. Lett. B} \textbf{\bibinfo{volume}{790}},
  \bibinfo{pages}{361} (\bibinfo{year}{2019}), \eprint{1812.03959}.

\bibitem[{\citenamefont{Boussarie et~al.}(2020)\citenamefont{Boussarie, Hatta,
  Szymanowski, and Wallon}}]{Boussarie:2019vmk}
\bibinfo{author}{\bibfnamefont{R.}~\bibnamefont{Boussarie}},
  \bibinfo{author}{\bibfnamefont{Y.}~\bibnamefont{Hatta}},
  \bibinfo{author}{\bibfnamefont{L.}~\bibnamefont{Szymanowski}},
  \bibnamefont{and} \bibinfo{author}{\bibfnamefont{S.}~\bibnamefont{Wallon}},
  \bibinfo{journal}{Phys. Rev. Lett.} \textbf{\bibinfo{volume}{124}},
  \bibinfo{pages}{172501} (\bibinfo{year}{2020}), \eprint{1912.08182}.

\bibitem[{\citenamefont{Hagiwara et~al.}(2020)\citenamefont{Hagiwara, Hatta,
  Pasechnik, and Zhou}}]{Hagiwara:2020mqb}
\bibinfo{author}{\bibfnamefont{Y.}~\bibnamefont{Hagiwara}},
  \bibinfo{author}{\bibfnamefont{Y.}~\bibnamefont{Hatta}},
  \bibinfo{author}{\bibfnamefont{R.}~\bibnamefont{Pasechnik}},
  \bibnamefont{and} \bibinfo{author}{\bibfnamefont{J.}~\bibnamefont{Zhou}},
  \bibinfo{journal}{Eur. Phys. J. C} \textbf{\bibinfo{volume}{80}},
  \bibinfo{pages}{427} (\bibinfo{year}{2020}), \eprint{2003.03680}.

\bibitem[{\citenamefont{Kovchegov and Santiago}(2021)}]{Kovchegov:2021iyc}
\bibinfo{author}{\bibfnamefont{Y.~V.} \bibnamefont{Kovchegov}}
  \bibnamefont{and} \bibinfo{author}{\bibfnamefont{M.~G.}
  \bibnamefont{Santiago}}, \bibinfo{journal}{JHEP}
  \textbf{\bibinfo{volume}{11}}, \bibinfo{pages}{200} (\bibinfo{year}{2021}),
  \eprint{2108.03667}.

\bibitem[{\citenamefont{Boer et~al.}(2022)\citenamefont{Boer, Hagiwara, Zhou,
  and Zhou}}]{Boer:2022njw}
\bibinfo{author}{\bibfnamefont{D.}~\bibnamefont{Boer}},
  \bibinfo{author}{\bibfnamefont{Y.}~\bibnamefont{Hagiwara}},
  \bibinfo{author}{\bibfnamefont{J.}~\bibnamefont{Zhou}}, \bibnamefont{and}
  \bibinfo{author}{\bibfnamefont{Y.-j.} \bibnamefont{Zhou}}
  (\bibinfo{year}{2022}), \eprint{2203.00267}.

\bibitem[{\citenamefont{Hatta et~al.}(2016)\citenamefont{Hatta, Xiao, and
  Yuan}}]{Hatta:2016dxp}
\bibinfo{author}{\bibfnamefont{Y.}~\bibnamefont{Hatta}},
  \bibinfo{author}{\bibfnamefont{B.-W.} \bibnamefont{Xiao}}, \bibnamefont{and}
  \bibinfo{author}{\bibfnamefont{F.}~\bibnamefont{Yuan}},
  \bibinfo{journal}{Phys. Rev. Lett.} \textbf{\bibinfo{volume}{116}},
  \bibinfo{pages}{202301} (\bibinfo{year}{2016}), \eprint{1601.01585}.

\bibitem[{\citenamefont{Balitsky}(1996)}]{Balitsky:1995ub}
\bibinfo{author}{\bibfnamefont{I.}~\bibnamefont{Balitsky}},
  \bibinfo{journal}{Nucl. Phys. B} \textbf{\bibinfo{volume}{463}},
  \bibinfo{pages}{99} (\bibinfo{year}{1996}), \eprint{hep-ph/9509348}.

\bibitem[{\citenamefont{Meissner et~al.}(2009)\citenamefont{Meissner, Metz, and
  Schlegel}}]{Meissner:2009ww}
\bibinfo{author}{\bibfnamefont{S.}~\bibnamefont{Meissner}},
  \bibinfo{author}{\bibfnamefont{A.}~\bibnamefont{Metz}}, \bibnamefont{and}
  \bibinfo{author}{\bibfnamefont{M.}~\bibnamefont{Schlegel}},
  \bibinfo{journal}{JHEP} \textbf{\bibinfo{volume}{08}}, \bibinfo{pages}{056}
  (\bibinfo{year}{2009}), \eprint{0906.5323}.

\bibitem[{\citenamefont{Collins et~al.}(1985)\citenamefont{Collins, Soper, and
  Sterman}}]{Collins:1984kg}
\bibinfo{author}{\bibfnamefont{J.~C.} \bibnamefont{Collins}},
  \bibinfo{author}{\bibfnamefont{D.~E.} \bibnamefont{Soper}}, \bibnamefont{and}
  \bibinfo{author}{\bibfnamefont{G.~F.} \bibnamefont{Sterman}},
  \bibinfo{journal}{Nucl. Phys. B} \textbf{\bibinfo{volume}{250}},
  \bibinfo{pages}{199} (\bibinfo{year}{1985}).

\bibitem[{\citenamefont{Xiao et~al.}(2017)\citenamefont{Xiao, Yuan, and
  Zhou}}]{Xiao:2017yya}
\bibinfo{author}{\bibfnamefont{B.-W.} \bibnamefont{Xiao}},
  \bibinfo{author}{\bibfnamefont{F.}~\bibnamefont{Yuan}}, \bibnamefont{and}
  \bibinfo{author}{\bibfnamefont{J.}~\bibnamefont{Zhou}},
  \bibinfo{journal}{Nucl. Phys. B} \textbf{\bibinfo{volume}{921}},
  \bibinfo{pages}{104} (\bibinfo{year}{2017}), \eprint{1703.06163}.

\bibitem[{\citenamefont{Zhou}(2019)}]{Zhou:2018lfq}
\bibinfo{author}{\bibfnamefont{J.}~\bibnamefont{Zhou}}, \bibinfo{journal}{Phys.
  Rev. D} \textbf{\bibinfo{volume}{99}}, \bibinfo{pages}{054026}
  (\bibinfo{year}{2019}), \eprint{1807.00506}.

\bibitem[{\citenamefont{Marquet and Soyez}(2005)}]{Marquet:2005zf}
\bibinfo{author}{\bibfnamefont{C.}~\bibnamefont{Marquet}} \bibnamefont{and}
  \bibinfo{author}{\bibfnamefont{G.}~\bibnamefont{Soyez}},
  \bibinfo{journal}{Nucl. Phys. A} \textbf{\bibinfo{volume}{760}},
  \bibinfo{pages}{208} (\bibinfo{year}{2005}), \eprint{hep-ph/0504080}.

\bibitem[{\citenamefont{Kovchegov}(1999)}]{Kovchegov:1999yj}
\bibinfo{author}{\bibfnamefont{Y.~V.} \bibnamefont{Kovchegov}},
  \bibinfo{journal}{Phys. Rev. D} \textbf{\bibinfo{volume}{60}},
  \bibinfo{pages}{034008} (\bibinfo{year}{1999}), \eprint{hep-ph/9901281}.

\bibitem[{\citenamefont{Bartels et~al.}(2000)\citenamefont{Bartels, Lipatov,
  and Vacca}}]{Bartels:1999yt}
\bibinfo{author}{\bibfnamefont{J.}~\bibnamefont{Bartels}},
  \bibinfo{author}{\bibfnamefont{L.~N.} \bibnamefont{Lipatov}},
  \bibnamefont{and} \bibinfo{author}{\bibfnamefont{G.~P.} \bibnamefont{Vacca}},
  \bibinfo{journal}{Phys. Lett. B} \textbf{\bibinfo{volume}{477}},
  \bibinfo{pages}{178} (\bibinfo{year}{2000}), \eprint{hep-ph/9912423}.

\bibitem[{\citenamefont{Lappi et~al.}(2016)\citenamefont{Lappi, Ramnath,
  Rummukainen, and Weigert}}]{Lappi:2016gqe}
\bibinfo{author}{\bibfnamefont{T.}~\bibnamefont{Lappi}},
  \bibinfo{author}{\bibfnamefont{A.}~\bibnamefont{Ramnath}},
  \bibinfo{author}{\bibfnamefont{K.}~\bibnamefont{Rummukainen}},
  \bibnamefont{and} \bibinfo{author}{\bibfnamefont{H.}~\bibnamefont{Weigert}},
  \bibinfo{journal}{Phys. Rev. D} \textbf{\bibinfo{volume}{94}},
  \bibinfo{pages}{054014} (\bibinfo{year}{2016}), \eprint{1606.00551}.

\bibitem[{\citenamefont{Contreras et~al.}(2020)\citenamefont{Contreras, Levin,
  Meneses, and Sanhueza}}]{Contreras:2020lrh}
\bibinfo{author}{\bibfnamefont{C.}~\bibnamefont{Contreras}},
  \bibinfo{author}{\bibfnamefont{E.}~\bibnamefont{Levin}},
  \bibinfo{author}{\bibfnamefont{R.}~\bibnamefont{Meneses}}, \bibnamefont{and}
  \bibinfo{author}{\bibfnamefont{M.}~\bibnamefont{Sanhueza}},
  \bibinfo{journal}{Phys. Rev. D} \textbf{\bibinfo{volume}{101}},
  \bibinfo{pages}{096019} (\bibinfo{year}{2020}), \eprint{2004.04445}.

\bibitem[{\citenamefont{McLerran and Venugopalan}(1994)}]{McLerran:1993ni}
\bibinfo{author}{\bibfnamefont{L.~D.} \bibnamefont{McLerran}} \bibnamefont{and}
  \bibinfo{author}{\bibfnamefont{R.}~\bibnamefont{Venugopalan}},
  \bibinfo{journal}{Phys. Rev. D} \textbf{\bibinfo{volume}{49}},
  \bibinfo{pages}{2233} (\bibinfo{year}{1994}), \eprint{hep-ph/9309289}.

\bibitem[{\citenamefont{Adloff et~al.}(2001)}]{H1:2001ert}
\bibinfo{author}{\bibfnamefont{C.}~\bibnamefont{Adloff}} \bibnamefont{et~al.}
  (\bibinfo{collaboration}{H1}), \bibinfo{journal}{Phys. Lett. B}
  \textbf{\bibinfo{volume}{520}}, \bibinfo{pages}{183} (\bibinfo{year}{2001}),
  \eprint{hep-ex/0108035}.

\bibitem[{\citenamefont{Koempel et~al.}(2012)\citenamefont{Koempel, Kroll,
  Metz, and Zhou}}]{Koempel:2011rc}
\bibinfo{author}{\bibfnamefont{J.}~\bibnamefont{Koempel}},
  \bibinfo{author}{\bibfnamefont{P.}~\bibnamefont{Kroll}},
  \bibinfo{author}{\bibfnamefont{A.}~\bibnamefont{Metz}}, \bibnamefont{and}
  \bibinfo{author}{\bibfnamefont{J.}~\bibnamefont{Zhou}},
  \bibinfo{journal}{Phys. Rev. D} \textbf{\bibinfo{volume}{85}},
  \bibinfo{pages}{051502} (\bibinfo{year}{2012}), \eprint{1112.1334}.

\bibitem[{\citenamefont{Lansberg et~al.}(2019)\citenamefont{Lansberg,
  Massacrier, Szymanowski, and Wagner}}]{Lansberg:2018fsy}
\bibinfo{author}{\bibfnamefont{J.~P.} \bibnamefont{Lansberg}},
  \bibinfo{author}{\bibfnamefont{L.}~\bibnamefont{Massacrier}},
  \bibinfo{author}{\bibfnamefont{L.}~\bibnamefont{Szymanowski}},
  \bibnamefont{and} \bibinfo{author}{\bibfnamefont{J.}~\bibnamefont{Wagner}},
  \bibinfo{journal}{Phys. Lett. B} \textbf{\bibinfo{volume}{793}},
  \bibinfo{pages}{33} (\bibinfo{year}{2019}), \eprint{1812.04553}.

\end{thebibliography}

\end{document}